\documentclass{aa}
\usepackage{graphicx} 
\begin{document}

%
%
\title{The optical emission line spectrum of Mark\,110
\thanks{Based on observations obtained with the Hobby-Eberly
               Telescope, which is a joint project of the University
               of Texas at Austin, the Pennsylvania State
           University, Stanford University, 
Ludwig-Maximilians-Universit\"at
            M\"unchen, and Georg-August-Universit\"at G\"ottingen.}}

\author{M.-P. V\'eron-Cetty \inst{1}, P. V\'eron \inst{1}, M. Joly \inst{2} and
W. Kollatschny \inst{3}}  

\offprints{P. V\'eron}

\institute
{Observatoire de Haute Provence, CNRS, F-04870 Saint-Michel l'Observatoire,
    France\\ 
\email{mira.veron@oamp.fr; philippe.veron@oamp.fr}
\and
Observatoire de Paris-Meudon, CNRS, Universit\'e Paris-Diderot, 5 place J. Janssen, F-92195 Meudon, France\\
\email{Monique.Joly@obspm.fr}
\and
Institute f\"{u}r Astrophysik, Universit\"{a}t G\"{o}ttingen, Friedrich-Hund-Platz 1, D-37077 
G\"{o}ttingen, Germany \\
\email{wkollat@astro.physik.uni-goettingen.de}}
\titlerunning{Mark\,110}
\authorrunning{V\'eron-Cetty et al.}

\date{Received ; accepted }

\abstract  
  {}
{We analyse in detail the rich emission line spectrum of Mark 110 to determine 
the physical conditions in the nucleus of this object, a peculiar NLS1 
without any detectable Fe II emission associated with the broad line region and with a 
$\lambda$5007/H$\beta$ line ratio unusually large for a NLS1.}
{We use 24 spectra obtained with the Marcario Low Resolution Spectrograph attached 
at the prime focus of the 9.2 m Hobby-Eberly telescope at the McDonald observatory. We fitted 
the spectrum by identifying all the emission lines (about 220) detected in the wavelength 
range 4200-6900 \AA\ (at rest).}
{The narrow emission lines are probably produced in a region with a density 
gradient in the range 10$^{3}$-10$^{6}$ cm$^{-3}$ with a rather high column density 
(5$\times$10$^{21}$ cm$^{-2}$). In addition to a narrow line system, three major broad line 
systems with different line velocity and width are required. We confirm the absence of broad 
Fe\,II emission lines. We speculate that Mark\,110 is in fact a BLS1 with relatively "narrow" 
broad lines but with a BH mass large enough compared to its luminosity to have a lower than 
Eddington luminosity.}
{}

\keywords{galaxies: Seyfert--galaxies: individual: Mark\,110}

\maketitle

\today


\section{Introduction}

 Narrow line Seyfert 1 galaxies (NLS1s) are Seyfert 1 galaxies in which the broad 
emission lines are relatively narrow
($<$2\,000 km s$^{-1}$ FWHM)(Osterbrock \& Pogge \cite{osterbrock85}). These objects 
generally have strong Fe\,II emission and relatively weak [O\,III]$\lambda$5007 emission
(Boroson \& Green \cite{boroson92}). However Grupe et al. (\cite{grupe99}; \cite{grupe04}) have found a few 
objects with "narrow" broad Balmer lines which have both weak Fe\,II emission and strong 
[O\,III]. Mark\,110, Kaz\,320 and HS\,0328+0528 are three such objects. 1RXS\,J102012.6+342837 
and 1RXS\,J133209.8+842412 could be two additional exemples (Wu et al. \cite{wu03}).

 The aim of this paper is to study the rich optical emission line spectrum of Mark\,110, 
one of these rare objects.

In section 2 we describe the target Mark\,110 and present the observations in section 3. In section 4
we analyse the emission line spectrum and in section 5 we determine the physical conditions in the NLR,
discuss the various determinations of the BH mass and the true nature of Mark\,110: NLS1 or BLS1. Our conclusions 
are summarized in section 6. 

\section{The target}

 Mark\,110 (0921+52) was discovered by Markaryan (\cite{markaryan69}) in the course of a 
slitless spectroscopic survey for UV excess galaxies. It was classified as a Seyfert 1 
(Arakelyan et al. \cite{arakelyan70}). The starlike object located 6$\arcsec$ to the north 
east of the nucleus is a star. The galaxy has a disturbed morphology suggestive of a recent 
merger (Wehinger \& Wyckoff \cite{wehinger77}; Adams \cite{adams77}; Hutchings \& Craven 
\cite{hutchings88}; McKenty \cite{mckenty90}; Bischoff \& Kollatschny \cite{bischoff99}). 
The Galactic extinction is 
A$_{V}$=0.056 mag. (Schlegel et al. \cite{schlegel98}). The redshift, measured from the 
[O\,III]$\lambda$5007 line is z=0.0352 (Vrtilek \& Carleton \cite{vrtilek85}). The H$\beta$ 
FWHM lies in the range 1\,670-2\,500 km s$^{-1}$ (Osterbrock \cite{osterbrock77}; Peterson 
et al. \cite{peterson85}; Crenshaw \cite{crenshaw86}; Boroson \& Green \cite{boroson92}; 
Bischoff \& Kollatschny \cite{bischoff99}; Stepanian et al. \cite{stepanian03}; Grupe 
et al. \cite{grupe04b}). Bischoff \& Kollatschny (\cite{bischoff99}) and Grupe et al. 
(\cite{grupe04b}) classified it as an NLS1 on the basis of its H$\beta$ FWHM (1670$\pm$50 
and 1760$\pm$50 km s$^{-1}$ respectively) measured after removal of the narrow 
component. 

  The optical continuum of Mark\,110 is variable (Peterson et al. \cite{peterson84}; 
\cite{peterson98}) with possible intranight variability (Webb \& Malkan \cite{webb00}).
The broad emission lines show strong variability (Peterson et al. \cite{peterson85}; 
Peterson \cite{peterson88}). The r.m.s. spectrum clearly shows H$\alpha$, H$\beta$ and 
H$\gamma$, He\,II $\lambda$4686 and the He\,I $\lambda$4471, $\lambda$4922, 
$\lambda$5016, $\lambda$5876 and $\lambda$6678 lines. The [Fe\,X] $\lambda$6375 line is also 
variable (Kollatschny et al. \cite{kollatschny01}). 

 The He\,II $\lambda$4686 line shows the largest variation of nearly a factor of 8 within 
two years. On the other hand H$\beta$ and the continuum at 5100 \AA\ vary only by 
a factor of 1.7 and 3.0 respectively within the same time interval (Bischoff \& Kollatschny 
\cite{bischoff99}; Peterson et al. \cite{peterson98}; \cite{peterson04}).

 There is a very broad component ($\sim$5\,000 km s$^{-1}$ FWHM), redshifted by 400$\pm$100 
km s$^{-1}$ with respect to the narrow lines, visible in the Balmer line profiles especially 
when the continuum is strong. This very broad component is the strongest contributor to the 
He\,II variability (Bischoff \& Kollatschny \cite{bischoff99}). The outer wings of the line 
profiles respond much faster to continuum variations than the central regions (Kollatschny 
\cite{kollatschny03a}).

 The Fe\,II emission is weak (the line ratio relative to H$\beta$ is R$_{4570}$=0.09-0.16) 
(Osterbrock \cite{osterbrock77}; Meyers \& Peterson \cite{meyers85}; Boroson \& Green 
\cite{boroson92}). The Fe\,II line flux remains constant while the Balmer line flux varies 
(Bischoff \& Kollatschny \cite{bischoff99}). 

\section{The observations}

 Twenty six spectra of Mark\,110 have been obtained between 1999, November 13 and
2000, May 14 with the Marcario Low Resolution Spectrograph (LRS) attached at the 
prime focus of the 9.2-m Hobby-Eberly telescope (HET) at McDonald observatory. 
The log of the observations is given in Table \ref{log}.
The detector was a 3072$\times$1024 15\,$\mu$m pixel Ford Aerospace CCD with 
2$\times$2 binning. The spectra cover the wavelength range 4200-6900\,\AA\ in the 
restframe of the galaxy, with a resolving power of 650 at 5000 \AA\ (7.7 \AA\ 
FWHM). Exposure times were 10 to 20 min.
  The slit width was 2$\farcs$0 ({\it i.e.} 75 $\mu$m or 3 pixels on the detector). 
Seven columns were extracted, corresponding to 3$\farcs$3 on the sky. Observations 
of several spectrophotometric standard stars were obtained to allow flux calibration 
of the spectra which have not been corrected for atmospheric absorption. Wavelength 
calibration was achieved via observations of HgCdZn and Ne spectra (Kollatschny et al. 
\cite{kollatschny01}).
Two of the spectra (2000 February 21 and April 30) of lower quality were ignored.
All spectra were deredshifted using z=0.0355 \footnote {Throughout this paper, we assume
 H$_{\rm o}$= 70 km s$^{-1}$ Mpc$^{-1}$.}.

We give in cols. 3 and 4 of Table \ref{log} the continuum flux 
in the wavelength range 5130-5140 \AA\ as measured by Kollatschny et al. (\cite 
{kollatschny01}) and the continuum flux at 5100 \AA\ as obtained from our fit, {\it 
i.e.} the value at 5100 \AA\ of the polynomial used for the continuum in the 
simultaneous fit of all emission lines in each individual spectrum after subtraction 
of an elliptical template (see below). In principle the difference between these two 
sets of numbers should be constant. It is not the case because of the different procedures 
used.
 

\begin{table}[ht]
\caption{\label{log} Log of observations. Col. 1: Julian date-2\,400\,000, 
col. 2: UT date, col. 3: continuum fluxes at 5100 \AA\ measured by Kollatschny 
et al. (\cite{kollatschny01}), col. 4: our measurements of the continuum flux
(in unit of 10$^{-15}$ erg$^{-1}$ cm$^{-2}$ \AA$^{-1}$) after removal of the 
host galaxy contribution.}
\begin{center} 
\begin{tabular}{|c|c|c|c|}
\hline
 JD & UT date & col. 3 & col. 4 \\
\hline
51495.94 & 1999.11.13 & 1.54 & 1.26 \\
51497.91 & 1999.11.15 & 1.56 & 1.25 \\
51500.91 & 1999.11.18 & 1.65 & 1.34 \\
51518.89 & 1999.12.06 & 1.92 & 1.51 \\
51520.87 & 1999.12.08 & 1.92 & 1.53 \\
51522.88 & 1999.12.10 & 1.94 & 1.53 \\
51525.84 & 1999.12.13 & 1.82 & 1.46 \\
51528.84 & 1999.12.16 & 1.86 & 1.49 \\
51547.80 & 2000.01.04 & 2.15 & 1.75 \\
51584.72 & 2000.02.10 & 1.41 & 1.10 \\
51586.71 & 2000.02.12 & 1.39 & 1.08 \\
51598.86 & 2000.02.24 & 1.63 & 1.30 \\
51605.83 & 2000.03.02 & 1.40 & 1.10 \\
51608.62 & 2000.03.05 & 1.35 & 1.07 \\
51611.62 & 2000.03.08 & 1.36 & 1.07 \\
51614.63 & 2000.03.11 & 1.09 & 0.80 \\
51629.76 & 2000.03.26 & 1.08 & 0.80 \\
51637.77 & 2000.04.03 & 1.04 & 0.77 \\
51645.73 & 2000.04.11 & 1.16 & 0.89 \\
51658.70 & 2000.04.24 & 1.38 & 1.12 \\
51663.68 & 2000.04.29 & 1.26 & 0.96 \\
51670.70 & 2000.05.06 & 1.33 & 1.01 \\
51673.69 & 2000.05.09 & 1.11 & 0.85 \\
51678.64 & 2000.05.14 & 1.11 & 0.82 \\
\hline
\end{tabular}
\end{center}
\end{table}

\section{Analysis}
 
 Using an HST image of Mark\,110, Bentz et al. (\cite{bentz06}) have shown that the 
contribution of the host galaxy in a 5$\farcs$0$\times$7$\farcs$6 aperture is equal
to 1.11$\times$10$^{-15}$ erg s$^{-1}$ cm$^{-2}$ \AA$^{-1}$. We have measured 
the contribution of this galaxy in the 2$\farcs$0$\times$3$\farcs$3 aperture used here 
on the HST image (taken through a filter centered at 5580 \AA) kindly provided to us 
by M.C. Bentz to be 55\% of this value, {\it i.e.} 0.61$\times$10$^{-15}$ erg s$^{-1}$ 
cm$^{-2}$ \AA$^{-1}$. By trial and error, we  
estimated the contribution of the host galaxy, assumed to be an E galaxy, to be 
0.25$\times$10$^{-15}$ erg s$^{-1}$cm$^{-2}$ \AA$^{-1}$ in our entrance aperture at 
5100 \AA\ . This is significantly smaller than the value inferred from the HST image. 
It could be due to the fact that the host galaxy is of a latter type with shallower 
absorption lines. The assumption that the host is an E galaxy is justified 
by the fact that Bentz et al. (\cite{bentz06}) have obtained a good fit of the image by 
using a central PSF and a de Vaucouleurs profile. We have subtracted from all spectra 
the spectrum of an E galaxy with our estimated flux density. 

\begin{figure*}[ht]

\resizebox{17.5cm}{!}{\includegraphics{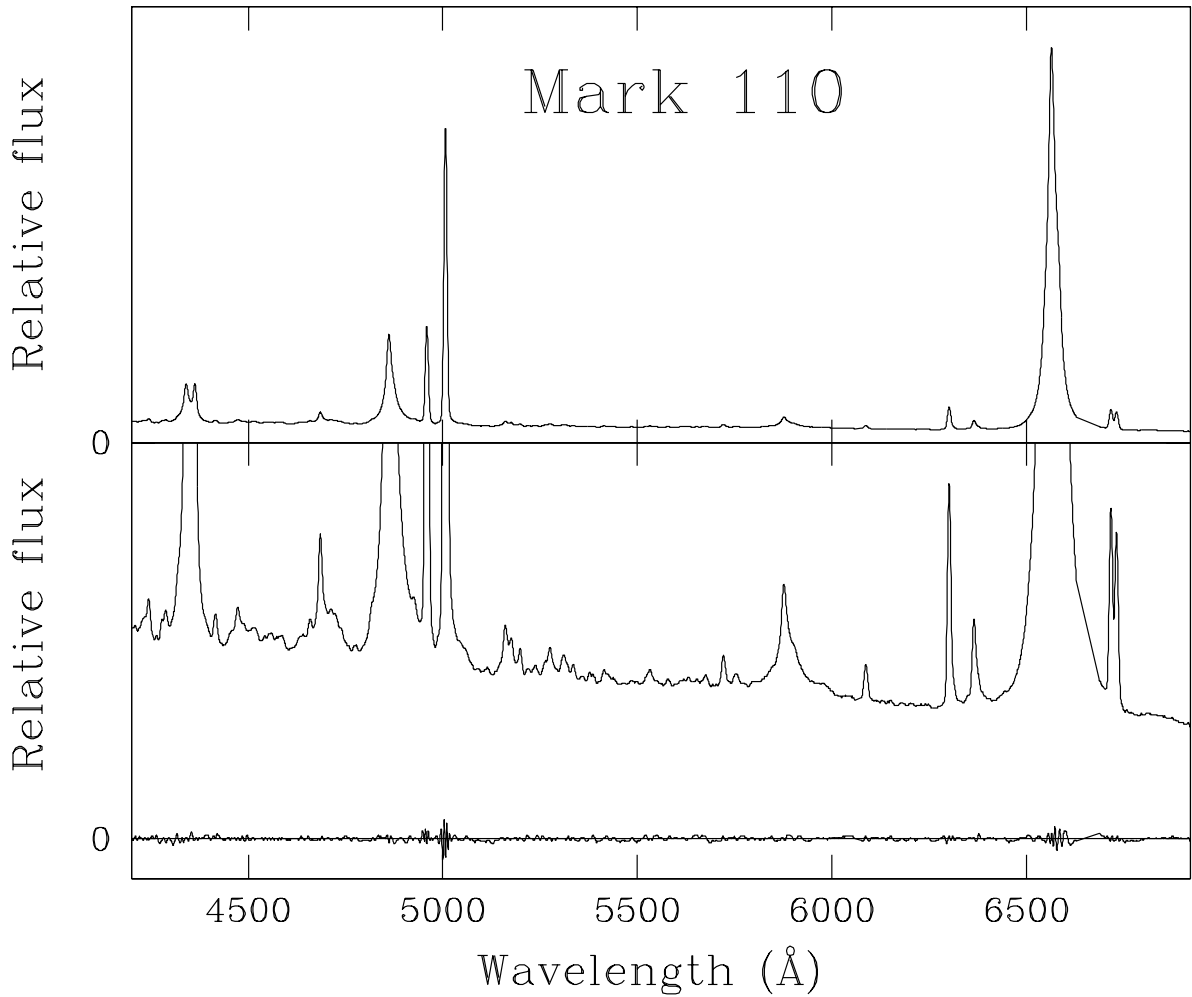}}

\caption{\label{Mark110_SP} The mean deredshifted HET spectrum of Mark\,110. The upper 
panel shows the spectrum. In the lower panel it is scaled to show the 
weak emission features. The residuals of the best fit are also shown.}

\end{figure*}

 We have averaged the 24 high quality HET spectra. This mean spectrum is shown 
in Fig.\,\ref{Mark110_SP}. To identify all individual emission lines and achieve a good 
fit of this spectrum\footnote{The fits were performed using a software originally 
written by E. Zuiderwijk and described in V\'eron et al. (\cite{veron80}).}, 
in addition to a narrow line system, three major line systems with different 
line velocity and width are required. 

\subsection{The narrow line system} 

 	Two components were needed to fit the strong narrow lines. The second, fainter, 
system is redshifted with respect to the first by 220 km s$^{-1}$. The H$\beta$ line ratio 
of these two systems is 0.20. 

The intrinsic [O\,III]$\lambda$5007 FWHM is equal to 280$\pm$3 km s$^{-1}$ (Feldman et al. 
\cite{feldman82}) or 288$\pm$5 km s$^{-1}$ (Vrtilek \& Carleton \cite{vrtilek85}). 
The resolution of our spectra is 475 km s$^{-1}$; 
we should therefore measure a FWHM of 550 km s$^{-1}$. 
The measured FWHM of the two components of [O\,III]\,$\lambda$5007 are equal to 520 and 470 
km s$^{-1}$ respectively.

  The [O\,III] $\lambda$5007 flux density has been measured to be equal to 
(2.26$\pm$0.14)$\times$10$^{-13}$ erg s$^{-1}$ cm$^{-2}$ by Peterson et al. 
(\cite{peterson98}). The spectra we used in this paper have been calibrated by
Bischoff \& Kollatschny (\cite{bischoff99}) in such a way that the [O\,III] 
$\lambda$5007 flux density is equal to this value. Adding the flux of the two
components needed in our model to fit this line, we obtain 2.19 $\times$10$^{-13}$ 
erg s$^{-1}$ cm$^{-2}$.

 The lines observed in the stonger system are listed in Table\,\ref{narrow}. They include 
lines of highly ionised ions ([Fe\,VI], [Fe\,VII], [Fe\,X], [Ca\,V], [Ca\,VII] which 
are slightly resolved (640 km s$^{-1}$ measured FWHM) and redshifted by $\sim$80 
km s$^{-1}$ with respect to the Balmer lines. De Robertis \& Osterbrock (\cite{robertis84}) 
and Appenzeller \& Ostreicher (\cite{appenzeller88}) have observed high-ionization lines 
in the emission spectrum of some Seyfert galaxies. In these objects, the very high 
ionization lines, especially those of [Fe\,VII], [Fe\,X] have FWHM which are broader that 
the typical low-ionization lines such as [O\,I] or [N\,I].

 The other lines, mostly from permitted and forbidden Fe\,II and Ti\,II, are redshifted 
by 100 km s$^{-1}$ with respect to the [O\,III] lines. The observed permitted Fe\,II
multiplets are listed in Table\,\ref{tfe2}. We have identified a line observed at $\sim$4480 
\AA\ with Mg II 4 $\lambda$4481. This line has been observed in emission in the eclipsing 
dwarf nova IP Peg (Harlaftis \cite{harlaftis99}) and in the "iron star" XX\,Oph (Merrill 
\cite{merrill61}; Cool et al. \cite{cool05}). 


\begin{table}[ht]
\caption{\label{tfe2} Observed permitted Fe\,II multiplets in the narrow line
system of the spectrum of Mark\,110. Col. 1: multiplet number, col. 2: transition, 
col. 3: upper level energy, col. 4: number of observed lines/number of lines in the 
multiplet in the observed spectral range.}
\begin{center}
\begin{tabular}{|r|c|c|c|}
\hline
 m. & Transition & u.l.(eV) &  \\
\hline
  42 & a$^{6}$S-z$^{6}$P$^{\rm o}$ & 5.34 & 3/3  \\
\hline
  27 & b$^{4}$P-z$^{4}$D$^{\rm o}$ & 5.56 & 1/6  \\
  38 & b$^{4}$F-z$^{4}$D$^{\rm o}$ & 5.56 & 6/9  \\
  43 & a$^{6}$S-z$^{4}$D$^{\rm o}$ & 5.56 & 1/3  \\
\hline
  37 & b$^{4}$F-z$^{4}$F$^{\rm o}$ & 5.57 & 6/10 \\
  49 & a$^{4}$G-z$^{4}$F$^{\rm o}$ & 5.57 & 6/9  \\
  55 & b$^{2}$H-z$^{4}$F$^{\rm o}$ & 5.57 & 1/3  \\
\hline
\end{tabular}
\end{center}
\end{table}

 This system shares many similarities with the emission line spectrum of the symbiotic 
nova RR\,Tel (McKenna et al. \cite{mckenna97}; Crawford et al. \cite{crawford99}) and 
that of XX\,Oph. The lines in XX\,Oph consist primarily of hydrogen and ionized metals 
such as Fe\,II, Cr\,II and Ti\,II.  Collin \& Joly (\cite{collin00}) have noted
that several types of stars, such as cataclysmic binaries, display intense Fe\,II lines;
it is believed that these lines are formed in the accretion disk. They suggested that 
physical conditions leading to their formation are similar to those in NLS1s. The fact 
that lines of Fe\,II, Cr\,II and Ti\,II are observed in some of these stars make their 
presence in the spectrum of Mark\,110 more plausible.\\

 In the weaker system, we observed the Balmer lines, He\,I $\lambda$5876, and the 
lines of [O\,I], [O\,III], [N\,II] and [S\,II]. \\

 The [O III]\,$\lambda$5007/H$\beta$ ratios are equal to 8.96 and 9.14 in the two systems 
respectively, while the [N\,II]\,$\lambda$6583/H$\alpha$ ratios are equal to 0.14 and 0.11 
and the [O\,I]\,$\lambda$6300/H$\alpha$ ratios are both equal to 0.20 (the Balmer 
line fluxes used in computing these line ratios are those of the relevant narrow components).
 The [O\,III] $\lambda$5007/$\lambda$4959 and [N\,II] $\lambda$6583/$\lambda$6548 
ratios have been set to their theoretical values of 3.01 and 3.07 respectively (Storey 
\& Zeippen \cite{storey00}). 
 The [O\,III] $\lambda$4363/$\lambda$5007 ratio R was measured to be equal to 0.086 
and 0.087 in the two regions. Osterbrock (\cite{osterbrock77}) measured R=0.039; 
this difference is unexplained. Our values suggest that the density in these 
regions is at least equal to 10$^{6.5}$ cm$^{-3}$ (Baskin \& Laor \cite{baskin05}). 
 The [N\,II] line ratio $\lambda$5754/($\lambda$6548+$\lambda$6584) is equal to 
0.057 in the strongest region which, for an electronic temperature of 10$^{4}$ K, 
would correspond to a density N$_{e}$$\sim$3$\times$10$^{5}$ cm$^{-3}$ (Keenan 
et al. \cite{keenan01}).
 The [S\,II]$\lambda$6716/$\lambda$6730 ratios are equal to 1.13 and 1.05 respectively,
suggesting that the density in the regions emitting these lines is of the order of
500$\times$(T$_{e}$/10$^{4}$)$^{0.5}$ cm$^{-3}$ (Osterbrock \cite{osterbrock74}). This
value is much smaller than the  one obtained from the [O\,III] lines indicating the 
presence of a density gradient among these clouds. The [Fe\,II] spectrum should arise 
in regions with N$_{e}$$<$10$^{6}$ cm$^{-3}$, otherwise these lines would be collisionally 
de-excited (Garcia-Lario et al. \cite{lario99}). \\

 These two NLR have almost identical spectra. They could perhaps be considered as two 
clouds belonging to a single entity, the strongest one being blueshifted with respect to 
the (unknown) systemic velocity of the galaxy by 110 km s$^{-1}$, the weakest one being 
redshifted by the same amount.
 
\subsection{The broad line systems}

 1/ A very broad line system, B1 ($\sim$ 6\,000 km s$^{-1}$ FWHM), is redshifted by 
$\sim$ 700 km s$^{-1}$ with respect to the strong narrow line system. The lines detected 
are H$\alpha$, H$\beta$, H$\gamma$, He\,II $\lambda$4686 and He\,I $\lambda$5876 and
$\lambda$6678. It can be identified with the very broad line system observed by Bischoff 
\& Kollatschny (\cite{bischoff99}), although they found a smaller line width ($\sim$ 5\,000 
km s$^{-1}$ FWHM); but the determination of the parameters of this system is made difficult 
by the presence of the atmospheric B band in the red wing of H$\alpha$, and therefore these two 
values may not be significantly different.

 2/ A broad line system, B2 (3\,340 km s$^{-1}$ FWHM), is redshifted by 440 km s$^{-1}$ 
with respect to the narrow line system. In this system, the only lines observed are
the Balmer lines (H$\alpha$, H$\beta$ and H$\gamma$). The Balmer decrement is 
H$\alpha$/H$\beta$=5.17.

 3/ A narrower line system, B3 (1\,515 km s$^{-1}$ FWHM), is redshifted by 180 km s$^{-1}$. 
In this system we found, in addition to the Balmer lines (H$\alpha$, H$\beta$ and H$\gamma$), 
He\,I lines ($\lambda$4471, $\lambda$4712, $\lambda$4922, $\lambda$5016 and $\lambda$5876) 
and He\,II $\lambda$4686.

 We have also detected in this system the Si\,II lines $\lambda5041$, $\lambda5056$,
$\lambda5958$ and $\lambda5979$. 
All spectra show a bump in the red wing of the complex emission region around 
$\lambda$5871 which we have identified with the Si\,II 4 doublet $\lambda\lambda$5958,5979.
There is a strong red shoulder on the red side of the [O\,III]$\lambda$5007 line which 
has been attributed by Kollatschny et al. (\cite{kollatschny01}) to He\,I $\lambda$5016; 
this attribution however does not seems to be appropriate as this would imply for this line 
a large red shift which is not observed in any of the line systems. We suggest that this
shoulder is due to the Si\,II 5 triplet $\lambda\lambda$5041,5056.0,5056.3.
 Si\,II lines are expected in objects with strong Fe\,II emission (Phillips \cite{phillips78}),
however we have not been able to detect any Fe\,II lines associated with this system; it is 
therefore rather surprising to observe these lines.

 Kollatschny et al. (\cite{kollatschny81}) found in the r.m.s. spectrum a variable line 
which they identified with [Fe\,X] $\lambda$6375. If this is the case, this line would be 
significantly broader than the other highly ionized Fe lines. We found this line to vary 
proportionally to H$\beta$. \\

 According to Bischoff \& Kollatschny (\cite{bischoff99}), all broad line profiles showed 
during the period 1987-1995 a red asymetry which would mainly be caused by a second line 
component redshifted by 1\,200 km s$^{-1}$. We found no evidence for such a component which 
may have been weak during the period studied here. 

\subsection{Variability of the broad emission lines} 

 Although it is difficult to observe the variability of Fe\,II, these lines seem to 
follow the variations of the continuum in a number of Seyfert 1s (Kollatschny \& Fricke 
\cite{kollatschny81}; Kollatschny et al. \cite{kollatschny81}; \cite{kollatschny00}; 
Vestergaard \& Peterson \cite{vestergaard06}; Wang et al. \cite{wang05}). In Mark\,110, 
the difference spectrum as well as the r.m.s. spectrum show no sign of variable Fe\,II 
emission. It seems therefore that there is no Fe\,II emission associated with the broad 
emission line region. 

\begin{figure}[ht]

\resizebox{8.5cm}{!}{\includegraphics{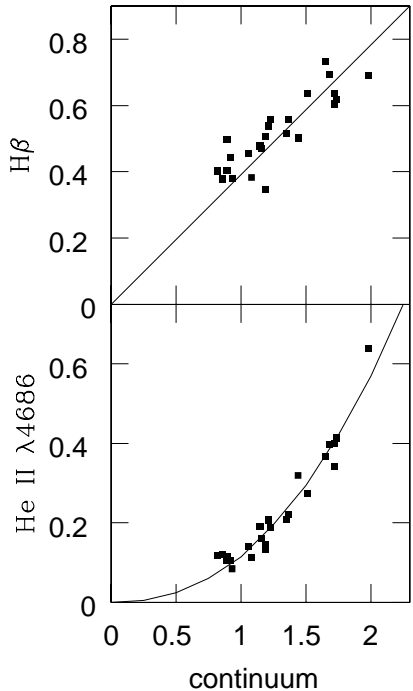}}

\caption{\label{hb4686} The upper panel shows the H$\beta$ flux of the very broad system
(B1) {\it vs} the continuum flux, while the lower panel shows the He\,II $\lambda$4686 
flux of this same system {\it vs} the continuum flux. In the lower panel, the curve shows 
the continuum at the power 2.3. The continuum fluxes are in units of 10$^{-15}$ erg 
s$^{-1}$ cm$^{-2}$ \AA$^{-1}$, while the line intensities are in units of 10$^{-15}$ erg 
s$^{-1}$ cm$^{-2}$.}

\end{figure}

 To study the variability of the broad emission lines, we have fitted all 24 individual 
spectra by setting the intensity of the narrow emission lines to the values found in the 
fit of the mean spectrum, keeping free only the intensities of the broad lines.

 The H$\beta$ intensity in the very broad line system (B1) is proportional to the continuum
intensity while He\,II $\lambda$4686 varies approximately as the power 2.3 of the continuum
intensity (Fig.\,\ref{hb4686}). This suggests that, when the continuum is bright, it is much 
bluer than when it is weak, as hydrogen is ionised by photons at 911 \AA\ while helium requires 
photons at 503 \AA.

 The H$\beta$ line of the narrower line system (B3) (1\,515 km s$^{-1}$ FWHM) varies 
significantly in the range (44-119)$\times$10$^{-15}$ erg s$^{-1}$cm$^{-2}$. The He\,I
$\lambda$5876 and He\,II $\lambda$4686 line intensities are proportional to H$\beta$ with
He\,I $\lambda$5876/H$\beta$=0.13 and He\,II $\lambda$4686/H$\beta$=0.08.

 The H$\beta$ line of the second system (B2) (3\,340 km s$^{-1}$ FWHM) varies with a much 
smaller amplitude if at all. When we set the intensities of the Balmer 
lines in this system to the values obtained from the mean spectrum, we achieve a good fit for 
all individual spectra.

\begin{figure}[ht]

\resizebox{8.8cm}{!}{\includegraphics{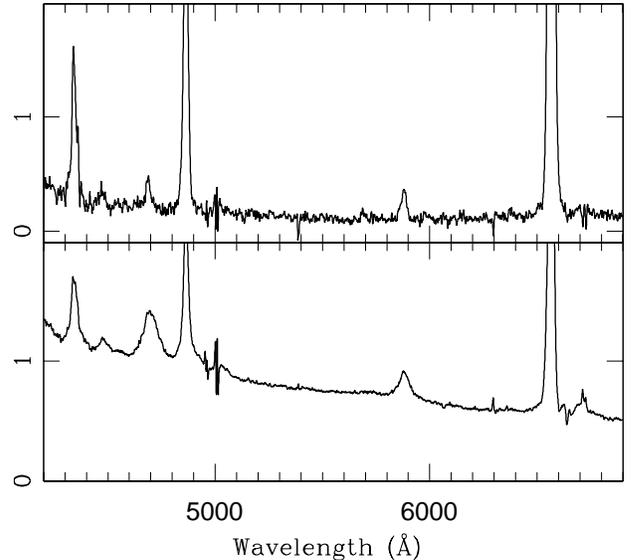}}

\caption{\label{mark110_var} The upper panel shows the difference between two spectra of 
Mark\,110 having the same continuum level. The very broad line system (B1) has completely 
disappeared. The only lines visible belong to the narrower broad line system (B3). The 
lower panel shows the difference between the mean spectrum of the four strongest spectra 
and the mean spectrum of the four weakest spectra.}
\end{figure}

 In Fig.\,\ref{mark110_var} (upper panel), we show the difference between two spectra of Mark\,110 
having almost the same continuum level (the difference between the mean of the two spectra 
of February 10 and 12, 2000 and the mean of the two spectra of April 29 and May 6, 2000). On this 
difference spectrum, all traces of the very broad lines (system B1) have disappeared, 
in agreement with the fact that these lines have a very small timelag (3.9$\pm$2.0 d) with respect 
to the continuum (Kollatschny \cite{kollatschny03b}). The velocity and FWHM of H$\alpha$ are 177
and 1295 km s$^{-1}$ respectively, very similar to the values found for component B3 (180 and 1515 
km s$^{-1}$). Component B2 shows no variation between the two epochs considered, separated by almost 
three months. The absence of variability of this component is most surprising.

 We have subtracted the mean of the four weaker spectra from the mean of the four stronger 
(Fig.\,\ref{mark110_var}, lower panel). The non variable lines disappear from the resulting 
spectrum. The remaining broad lines are those seen by Kollatschny et al. (\cite{kollatschny01}) 
on the r.m.s. spectrum.
   
\section{Discussion}

\subsection{The physical conditions in the NLR}

 From the line ratios given in section 4.1 we have an estimate of the density range of 
the emission regions producing the main forbidden emission lines: [O\,III], 
[N\,II], [S\,II]. The photoionization code CLOUDY (Ferland 2002) allows us to define 
more precisely the physical parameters of the medium responsible for the bulk of 
the emission lines detected in the NLR. Adopting a mean optical luminosity equal 
to 5$\times$10$^{43}$ ergs s$^{-1}$ and a power law slope of the ionizing radiation
$\alpha$=--1.0 at energies higher than 0.06 Ryd, we have calculated a number of 
models using the large Fe$^+$ atom to match the observed narrow Fe\,II lines in 
addition to the permitted and forbidden lines identified in the NLR. Abundances are 
about solar (C: --3.61; N: --4.59; O: --3.31; Ne: --4.00; Na: --5.67; Mg: --4.46; Si: --4.46; 
S: --4.74; Ar: --5.60; Ca: --5.64; Fe: --4.07). However, CLOUDY does not include 
optical permitted lines of Ti\,II, Cr\,II or Si\,II. Tables \ref{modelf} 
and \ref{modelp} list respectively the forbidden and permitted emission 
lines which are both observed in the NLR and computed in the code. The observed line 
ratios referred to H$\beta$ (the H$\beta$ flux is 20.3$\times$10$^{-15}$ erg s$^{-1}$
cm$^{-2}$) are given in the third column of the tables while the predicted ones from
two different models are displayed in columns 4 and 5. These two models define the 
range of parameters of the set of discrete clouds with different physical states 
constituting the NLR. The best fit is obtained for densities (n) in the range 
10$^3$-10$^6$ cm$^{-3}$, with a column density (N$_{\rm H}$) of respectively 
5$\times$10$^{19}$ and 5$\times$10$^{21}$ cm$^{-2}$. The ionization parameter is of 
the order of 10$^{-3}$ which implies a cloud distance to the central source of 
radiation of 30 and 2000\,pc (R=10$^{20}$ to 6$\times$10$^{21}$ cm) depending on the 
density. The temperature in the low density clouds is around 10\,000\,K, while inside 
the high density cloud whose optical thickness is higher there is a gradient of 
temperature from 17\,000\,K to 6\,000\,K.

\begin{table}[ht]
\caption{\label{modelf} Observed and computed line ratios, referred to the flux of the relevant 
H$\beta$ component (20.3$\times$10$^{-15}$ erg s$^{-1}$ cm$^{-2}$), of the 
forbidden lines detected in the NLR of Mrk110. R is the distance of the emitting cloud 
from the ionizing source in cm, n the density in cm$^{-3}$ and N$_{\rm H}$ the column
density in cm$^{-2}$.}
\begin{tabular}{|c|c|c|c|c|c|}
\hline
lines &$\lambda$& Obs.  & model     & model  \\
      &   (\AA) &       &R=$10^{20}$ &R=6.$10^{21}$\\
      &         &       &n=$10^{6}$  &n=$10^{3}$  \\
      &         &       &$N_H$=$5.10^{21}$ & $N_H$=$5.10^{19}$ \\
\hline
$[O\,I]$   & 5577   & 0.01 &0.02  & 0.00 \\
$[O\,I]$   & 6300   & 0.60 &1.30  & 0.01 \\
$[O\,I]$   & 6363   & 0.20 &0.42  & 0.00 \\
$[O\,III]$ & 4363   & 0.77 &0.81  & 0.08 \\
$[O\,III]$ & 4959   & 2.97 &2.93  & 3.54 \\
$[O\,III]$ & 5007   & 8.96 &8.83  &10.60 \\
           &        &      &      &      \\
$[N\,I]$   & 5198   & 0.01 &0.00  & 0.00 \\
$[N\,I]$   & 5200   & 0.04 &0.00  & 0.00 \\
$[N\,II]$  & 5755   & 0.03 &0.14  & 0.02 \\
$[N\,II]$  & 6548   & 0.14 &0.32  & 0.49 \\
$[N\,II]$  & 6584   & 0.43 &0.96  & 1.43 \\
           &        &      &      &      \\
$[S\,II]$  & 6716   & 0.58 &0.04  & 0.57 \\
$[S\,II]$  & 6731   & 0.51 &0.10  & 0.68 \\
$[S\,III]$ & 6312   & 0.07 &0.18  & 0.07 \\
           &        &      &      &      \\
$[Ar\,III]$&5192    & 0.03 & 0.01 & 0.00 \\
$[Ar\,IV]$ & 4711   & 0.01 & 0.00 & 0.01 \\
$[Ar\,IV]$ & 4740   & 0.04 & 0.03 & 0.01 \\
           &        &      &      &      \\
$[Ne\,IV]$ & 4720   & 0.08 & 0.03 & 0.00 \\
           &        &      &      &      \\
$[Ca\,V]$  & 5309   & 0.08 & 0.03 & 0.00 \\
$[Ca\,VII]$& 5620   & 0.01 & 0.0  & 0.00 \\
           &        &      &      &      \\
$[Fe\,II]$  4F  & 4639 & 0.00 &0.06 & 0.00 \\
$[Fe\,II]$  4F  & 4728 & 0.01 &0.13 & 0.00 \\
$[Fe\,II]$  4F  & 4798 & 0.00 &0.02 & 0.00 \\
$[Fe\,II]$  4F  & 4890 &      &0.19 & 0.01 \\
$[Fe\,II]$  6F  & 4416 & 0.05 &0.23 & 0.01 \\
$[Fe\,II]$  6F  & 4432 & 0.00 &0.02 & 0.00 \\
$[Fe\,II]$  6F  & 4458 & 0.02 &0.14 & 0.00 \\
$[Fe\,II]$  6F  & 4488 & 0.01 &0.07 & 0.00 \\
$[Fe\,II]$  6F  & 4493 & 0.01 &0.03 & 0.00 \\
$[Fe\,II]$  6F  & 4515 & 0.00 &0.03 & 0.00 \\
$[Fe\,II]$  6F  & 4528 & 0.00 &0.02 & 0.00 \\
$[Fe\,II]$  7F  & 4287 & 0.09 &0.25 & 0.01 \\
$[Fe\,II]$  7F  & 4359 & 0.07 &0.18 & 0.01 \\
$[Fe\,II]$  7F  & 4414 & 0.05 &0.13 & 0.01 \\
$[Fe\,II]$  7F  & 4452 & 0.03 &0.08 & 0.00 \\
$[Fe\,II]$  7F  & 4475 & 0.01 &0.04 & 0.00 \\
$[Fe\,II]$ 17F  & 5412 & 0.03 &0.05 & 0.00 \\
$[Fe\,II]$ 17F  & 5495 & 0.01 &0.03 & 0.00 \\
$[Fe\,II]$ 17F  & 5527 & 0.04 &0.12 & 0.00 \\
$[Fe\,II]$ 18F  & 5107 & 0.00 &0.04 & 0.00 \\
\hline
\end{tabular}
\end{table}
\addtocounter{table}{-1}
\begin{table}[ht]
\caption{ (continued)} 
\begin{tabular}{|c|c|c|c|c|c|}
\hline
lines &$\lambda$& Obs.   & model     & model  \\
      &  (\AA)  &        &R=$10^{20}$ &R=6.$10^{21}$\\
      &         &        &n=$10^{6}$  &n=$10^{3}$  \\
      &         &        &$N_H$=$5.10^{21}$ & $N_H$=$5.10^{19}$ \\
\hline
$[Fe\,II]$ 18F  & 5158 & 0.01 &0.10 & 0.00 \\
$[Fe\,II]$ 18F  & 5181 & 0.00 &0.05 & 0.00 \\
$[Fe\,II]$ 18F  & 5269 & 0.00 &0.06 & 0.00 \\
$[Fe\,II]$ 18F  & 5273 & 0.01 &0.25 & 0.01 \\
$[Fe\,II]$ 18F  & 5433 & 0.00 &0.08 & 0.00 \\
$[Fe\,II]$ 19F  & 5112 & 0.01 &0.10 & 0.00 \\
$[Fe\,II]$ 19F  & 5159 & 0.07 &0.52 & 0.04 \\
$[Fe\,II]$ 19F  & 5220 & 0.01 &0.10 & 0.00 \\
$[Fe\,II]$ 19F  & 5262 & 0.04 &0.33 & 0.02 \\
$[Fe\,II]$ 19F  & 5297 & 0.01 &0.07 & 0.00 \\
$[Fe\,II]$ 19F  & 5334 & 0.03 &0.24 & 0.00 \\
$[Fe\,II]$ 19F  & 5376 & 0.02 &0.20 & 0.00 \\
$[Fe\,II]$ 20F  & 4775 & 0.01 &0.07 & 0.00 \\
$[Fe\,II]$ 20F  & 4815 & 0.05 &0.23 & 0.01 \\
$[Fe\,II]$ 20F  & 4874 & 0.01 &0.09 & 0.00 \\
$[Fe\,II]$ 20F  & 4905 & 0.02 &0.12 & 0.00 \\
$[Fe\,II]$ 20F  & 4947 & 0.01 &0.03 & 0.00 \\
$[Fe\,II]$ 20F  & 4951 & 0.01 &0.07 & 0.00 \\
$[Fe\,II]$ 20F  & 4973 & 0.01 &0.07 & 0.00 \\
$[Fe\,II]$ 20F  & 5005 & 0.01 &0.04 & 0.00 \\
$[Fe\,II]$ 20F  & 5020 & 0.01 &0.07 & 0.00 \\
$[Fe\,II]$ 20F  & 5043 & 0.01 &0.04 & 0.00 \\
$[Fe\,II]$ 21F  & 4244 & 0.10 &0.23 & 0.01 \\
$[Fe\,II]$ 21F  & 4245 & 0.02 &0.06 & 0.00 \\
$[Fe\,II]$ 21F  & 4277 & 0.06 &0.16 & 0.01 \\
$[Fe\,II]$ 21F  & 4306 & 0.02 &0.05 & 0.00 \\
$[Fe\,II]$ 21F  & 4320 & 0.04 &0.11 & 0.00 \\
$[Fe\,II]$ 21F  & 4347 & 0.02 &0.05 & 0.00 \\
$[Fe\,II]$ 21F  & 4353 & 0.03 &0.07 & 0.00 \\
$[Fe\,II]$ 21F  & 4358 & 0.04 &0.11 & 0.00 \\
$[Fe\,II]$ 21F  & 4372 & 0.02 &0.05 & 0.00 \\
$[Fe\,II]$ 35F  & 5163 & 0.05 &0.05 & 0.00 \\
$[Fe\,II]$ 35F  & 5199 & 0.01 &0.02 & 0.00 \\
$[Fe\,II]$ 35F  & 5278 & 0.01 &0.01 & 0.00 \\
$[Fe\,II]$ 35F  & 5283 & 0.01 &0.01 & 0.00 \\
                &      &      &     &      \\
$[Fe\,III]$ 3F  & 4658 & 0.05 &0.37 & 0.75 \\
$[Fe\,III]$ 1F  & 4931 & 0.05 &0.04 & 0.02 \\
$[Fe\,III]$ 1F  & 5271 & 0.03 &0.21 & 0.41 \\
                &      &      &     &      \\
$[Fe\,VI]$  2F  & 5177 & 0.14 &0.10 & 0.01 \\
                &      &      &     &      \\
$[Fe\,VII]$ 2F  & 4894 & 0.01 &0.02 & 0.00 \\
$[Fe\,VII]$ 2F  & 4943 & 0.04 &0.03 & 0.00 \\
$[Fe\,VII]$ 2F  & 5159 & 0.09 &0.03 & 0.00 \\
$[Fe\,VII]$ 2F  & 5277 & 0.05 &0.03 & 0.00 \\
$[Fe\,VII]$ 1F  & 5721 & 0.13 &0.11 & 0.00 \\
$[Fe\,VII]$ 1F  & 6087 & 0.18 &0.17 & 0.00 \\
                &      &      &     &      \\
$[Fe\,X]$   1F  & 6373 & 0.01 &0.0  & 0.00 \\

\hline
\end{tabular}
\end{table}



\begin{table}[ht]
\caption{\label{modelp}Same as table 3 for the permitted lines.}
\begin{tabular}{|c|c|c|c|c|}
\hline
lines &$\lambda$& Obs.  & model     & model  \\
      &  (\AA)  &       &R=$10^{20}$ &R=6.$10^{21}$\\
      &         &      &n=$10^{6}$  &n=$10^{3}$  \\
      &         &      &$N_H$=$5.10^{21}$ & $N_H$=$5.10^{19}$ \\
\hline

H$\alpha$& 6563    & 3.02 &2.91 & 2.87 \\
H$\beta $& 4861    & 1.00 &1.00 & 1.00 \\
H$\gamma$& 4340    & 0.48 &0.47 & 0.47 \\
         &         &      &     &      \\
He\,II   & 4339    & 0.01 &0.01 & 0.01 \\
He\,II   & 4542    & 0.05 &0.01 & 0.01 \\
He\,II   & 4686    & 0.22 &0.26 & 0.37 \\
He\,II   & 5412    & 0.00 &0.02 & 0.03 \\
He\,II   & 6560    & 0.02 &0.04 & 0.05 \\
He\,I    & 4388    & 0.00 &0.00 & 0.00 \\
He\,I    & 4471    & 0.05 &0.04 & 0.03 \\
He\,I    & 4713    & 0.00 &0.01 & 0.00 \\
He\,I    & 4922    & 0.01 &0.01 & 0.01 \\
He\,I    & 5016    & 0.02 &0.02 & 0.01 \\
He\,I    & 5876    & 0.10 &0.13 & 0.08 \\
He\,I    & 6678    & 0.02 &0.03 & 0.02 \\
Na\,ID   & 5892    & 0.07 &0.02 & 0.00 \\
         &         &      &     &      \\
Fe\,II m27 & 4233 & 0.02  &0.01 & 0.00 \\
Fe\,II m37 & 4489 & 0.01  &0.00 & 0.00 \\
Fe\,II m37 & 4491 & 0.03  &0.00 & 0.00 \\
Fe\,II m38 & 4508 & 0.05  &0.00 & 0.00 \\
Fe\,II m37 & 4515 & 0.05  &0.00 & 0.00 \\
Fe\,II m37 & 4520 & 0.02  &0.00 & 0.00 \\
Fe\,II m38 & 4522 & 0.03  &0.00 & 0.00 \\
Fe\,II m38 & 4549 & 0.04  &0.00 & 0.00 \\
Fe\,II m37 & 4555 & 0.02  &0.00 & 0.00 \\
Fe\,II m38 & 4576 & 0.03  &0.00 & 0.00 \\
Fe\,II m37 & 4582 & 0.00  &0.00 & 0.00 \\
Fe\,II m38 & 4583 & 0.03  &0.00 & 0.00 \\
Fe\,II m37 & 4629 & 0.03  &0.00 & 0.00 \\
Fe\,II m42 & 4924 & 0.05  &0.03 & 0.00 \\
Fe\,II m42 & 5018 & 0.08  &0.03 & 0.00 \\
Fe\,II m42 & 5169 & 0.03  &0.05 & 0.00 \\
Fe\,II m49 & 5197 & 0.01  &0.00 & 0.00 \\
Fe\,II m49 & 5234 & 0.03  &0.00 & 0.00 \\
Fe\,II m49 & 5275 & 0.03  &0.00 & 0.00 \\
Fe\,II m49 & 5316 & 0.04  &0.01 & 0.00 \\
Fe\,II m49 & 5325 & 0.02  &0.00 & 0.00 \\
Fe\,II m49 & 5425 & 0.02  &0.00 & 0.00 \\

\hline
\end{tabular}
\end{table}

 The low density/low column density clouds partly account  for the
Balmer, He\,I and He\,II lines as well as for [O\,III]$\lambda\lambda$4959,
5007, [N\,II]$\lambda\lambda$6548,6584, and [S\,II]$\lambda\lambda$6716,
6731. The high density/high column density clouds account for the same
lines (except [S\,II]) plus [O\,III]$\lambda$4363 and the [O\,I] lines,
but also partly for the weak component of permitted Fe\,II lines and
some high ionization lines such as [Ar\,IV], [Ca\,V], [Fe\,VI] and
[Fe\,VII]. The main discrepancies between the observed and predicted
line ratios involve the [N\,II] and [Fe\,III] lines which are
predicted to be too strong. A lower abundance of nitrogen would improve 
the [N\,II]/H$\beta$ ratio.

\subsection{The black hole mass}

 To estimate the mass of the central BH, the assumption has to be made that the motion 
of the BLR clouds is  gravitationally dominated (Peterson \& Wandel \cite{peterson00}) 
which may not be the case (Krolik \cite{krolik01}). Then the BH mass is given by
M$_{\rm BH}$=V$^{2}$$\times$R/G where G is the gravitational constant, R the radius of 
the BLR and V the Keplerian velocity of the emitting cloud (Kaspi et al. \cite{kaspi00}). 

  Reverberation mapping studies made it possible to determine the size of the BLR in a 
number of type 1 AGN, which led to the discovery of a correlation between the radius 
of the region emitting the H$\beta$ line and the monochromatic luminosity at 5100 \AA . 
The BLR size scales with the rest frame luminosity as L$^{0.52\pm0.04}$ (Kaspi et al. 
\cite{kaspi00}; \cite{kaspi05}; Bentz et al. \cite{bentz06}). The radius of the BLR is 
either estimated directly from reverberation mapping or by using this correlation.
  
 V is taken to be equal to k$\times$FWHM. The numerical factor k depends on the structure, 
kinematics and orientation of the BLR and is often assumed to be equal to $\sqrt{3}$/2 
corresponding to an isotropic BLR with random orbital motion (Netzer \cite{netzer90}). 
Peterson et al. (\cite{peterson04}), normalizing the AGN M$_{\rm BH}$-$\sigma$$_{*}$ 
relationship to the M$_{\rm BH}$-$\sigma$$_{*}$ relationship for quiescent galaxies (Onken 
et al. \cite{onken04}), found k=1.26 which leads to a BH mass 1.8 times larger. 

 Thus for a given luminosity, NLS1s have a smaller BH mass than BLS1s as the BH mass scales 
as the square of the line width while the Eddington ratio, {\it i.e.} the ratio of the 
bolometric to the Eddington luminosity (assuming that 
L$_{\rm bol}$$\sim$10$\times$$\lambda$$\times$L$_{\lambda}$(5100\AA)) is larger, sometimes 
greater than one, as shown {\it e.g.} by Collin \& Kawaguchi (\cite{collin04}) 
\footnote{Elvis et al. 
(\cite{elvis99}) however have shown that the dispersion of the values of the Eddington ratio
for a given BH mass is at least equal to a factor of 2.}. \\

 Kollatschny et al. (\cite{kollatschny01}), comparing the observed profile variations with
model calculations of different velocity fields, concluded that the broad line region of 
Mark\,110 is an accretion disc, implying that the BH mass is given by 
M$_{\rm BH}$=1.5$\times$FWHM$^{2}$$\times$R/G (k=1.22). They measured the H$\beta$ FWHM on 
the r.m.s. spectrum to be 1515$\pm$100 km s$^{-1}$ and a time lag for H$\beta$ of 24.2$\pm$3.5 
days. They obtained M$_{\rm BH}$=(1.8$\pm$0.4$)\times$10$^{7}$ M$\odot$ in good agreement 
with the value given by Onken et al. (\cite{onken04}): M$_{\rm BH}$=(2.5$\pm$0.6$)\times$10$^{7}$ 
M$\odot$. \\

 The line width of the r.m.s. spectrum (1\,670 and 1\,515 km s$^{-1}$) measured by Wandel et al. 
(\cite{wandel99}) and Kollatschny et al. (\cite{kollatschny01}) shows that the
variable component is our component B3. \\

 The bulge velocity dispersion was measured to be 86$\pm$13 km s$^{-1}$ by Ferrarese et al. 
(\cite{ferrarese01}) which would correspond to a BH mass of (0.25$\pm$0.10)$\times$10$^{7}$
M$\odot$ (Ferrarese \& Merrit \cite{ferrarese00}; Merritt \& Ferrarese \cite{merritt01}) or 
(0.33$\pm$0.18)$\times$10$^{7}$ M$\odot$ (Greene \& Ho \cite{greene06}). The [O\,III] emission 
line width has been extensively used as a representation of the bulge velocity dispersion.
However the [O\,III] value typically overestimates the stellar velocity dispersion by as much as 
a factor of two in NLS1s (Botte et al. \cite{botte05}). For Mark\,110, the velocity dispersion of the 
[O\,III] line is $\sim$120 km s$^{-1}$ (see above) or 39\% larger than the stellar velocity dispersion. 
The virial mass of 
the BH is about 6 times larger than expected from its bulge velocity dispersion (Ferrarese et 
al. \cite{ferrarese01}; Onken et al. \cite{onken04}). However Barth et al. (\cite{barth05}) 
showed that the virial BH mass dispersion around the M$_{\rm BH}$-$\sigma$$_{*}$ relationship 
is approximately equal to a factor of 4. \\
 
\begin{figure*}[ht]

\resizebox{17.5cm}{!}{\includegraphics{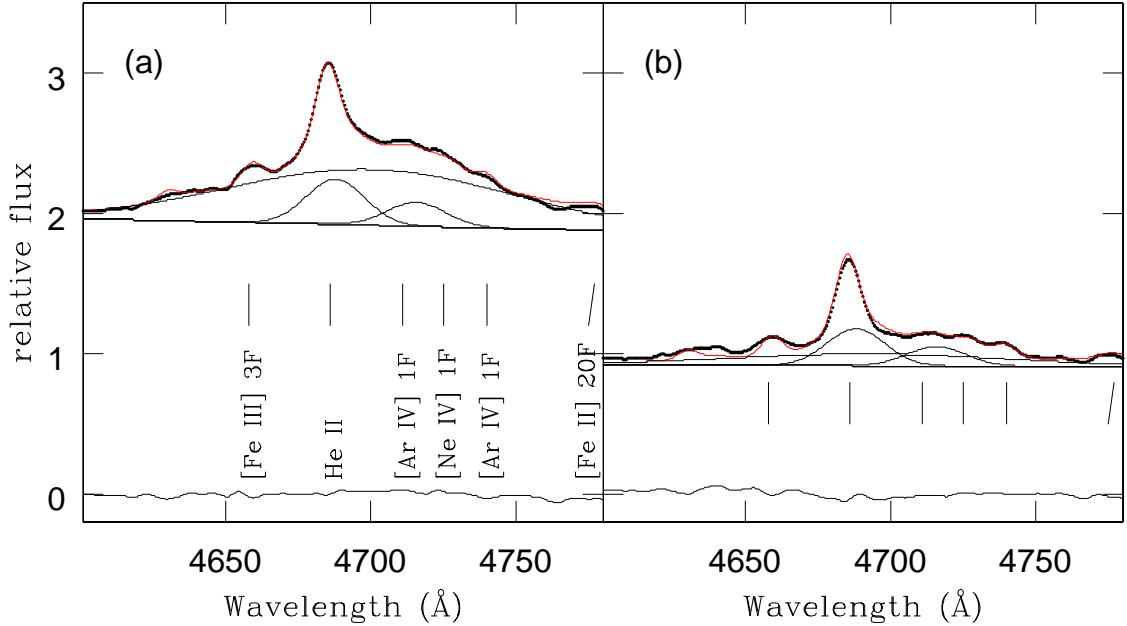}}

\caption{\label{HeII4686} (a) Fit of the mean of the four strongest spectra of Mark\,110 
in the region around He\,II 
$\lambda$4686 and He\,I $\lambda$4713. The dotted line shows the observations; the continuous 
line is the best fit. The lowest line represents the residuals. We also show the three He\,I and 
He\,II variable components. The locations of the strongest narrow lines are indicated. (b) The 
same for the mean of the four weakest spectra.}
\end{figure*}

 If the BLR is a rotating disk, the observed line width depends on its inclination to the 
line of sight. Collin \& Kawaguchi (\cite{collin04}) have shown that the BLR should be a 
geometrically thick disc. Such a disc must be sustained vertically 
by a turbulent pressure corresponding to a turbulent velocity which is such that, when seen 
face-on, the width of the emission lines emitted by the disc is reduced compared to an edge-on 
disc depending on the aspect ratio of the disc {\it h/r}. This could cause a systematic 
underestimation of the central mass by a factor of {\it (h/r)}$^{2}$ (Krolik \cite{krolik01}). 
 Accordingly, Kollatschny (\cite{kollatschny03b}) noted that the derived BH 
mass is a lower limit. He showed that the redshift of the r.m.s. profiles with respect to the 
narrow emission lines increases as a function of line width and ionization potential. He 
interpreted this effect as being due to gravitational redshifts. The BH mass needed to explain 
these redshifts is M$_{\rm BH}$=(14$\pm$3$)\times$10$^{7}$ M$\odot$, implying a value of the 
aspect ratio smaller than 0.36. We note however that, according to Kollatschny 
(\cite{kollatschny03b}), the broadest line, with the largest redshift, is He\,II 
$\lambda$4686. But the broad variable emission feature at $\sim$4700 \AA\ which was 
considered by Kollatschny as being a single very broad He\,II $\lambda$4686 component is 
modeled here, ignoring the non variable narrow lines, with three individual broad 
variable lines: He\,I $\lambda$ 4713 and He\,II $\lambda$4686 in the system B3 and He\,II 
$\lambda$4686 in the system B1 (Fig.\,\ref{HeII4686}). It turns out that the very broad 
He\,II line is not always the main contributor to the emission complex and is even barely 
detected in the 12 weakest spectra. In these conditions it seems doubtful that cross 
correlating the integrated flux of the broad emission feature with the intensity of the 
continuum can lead to a reliable timelag. 

 M\"{u}ller \& Wold (\cite{muller06}), modelling the emission lines emitted near a Kerr 
BH, have shown that the broad lines observed in Mark\,110 could indeed be gravitationally 
redshifted in an accretion disk having an inclination of $\sim$30\degr. Moreover, a broad 
($\sim$16\,200 km s$^{-1}$ FWHM), redshifted (z=0.023) component of the O\,VII triplet 
(at $\sim$570 eV) discovered in the spectrum of Mark\,110 could be due to a gravitational 
redshift effect; however, infall motion towards the central BH cannot be excluded (Boller 
et al. \cite{boller07}).

 This BH mass ((14$\pm$3$)\times$10$^{7}$ M$\odot$) is about 50 times larger than the value 
obtained from the bulge velocity dispersion which is unaffected by orientation effects but 
which could be influenced by the merging experienced by the host galaxy. It is also 
$\sim$7 times greater than the mass derived from reverberation mapping, but this difference 
could be explained, as we have seen, if the accretion disk is seen nearly pole-on with an 
aspect ratio smaller than 0.36.

 Papadakis (\cite{papadakis04}) has found a significant anticorrelation between the 2-10 
keV variability amplitude and the BH mass. The upper limit observed for the variance of 
Mark\,110 suggests that the BH mass is larger than 10$^{7}$ M$\odot$ (O'Neill et al. 
\cite{oneill05}), in agreement with the reverberation mapping determination.

\subsection{Nature of Mark\,110: NLS1 or BLS1?}

 Subtracting the narrow H$\beta$ components from the average spectrum of Mark\,110, we 
found that the broad emission line system has a width of $\sim$ 1\,700 km s$^{-1}$ (FWHM) 
and, therefore, this galaxy could be classified as a NLS1 as suggested by Grupe et al. 
(\cite{grupe04b}). However, this object has none of the other properties characteristic 
of NLS1s. \\

 NLS1s generally have a soft X-ray excess together with an unusually steep 2-10 keV 
power law which could be due to a high accretion rate (Pounds et al. \cite{pounds95};
Shemmer et al. \cite{shemmer06}), although strong ultra soft X-ray emission is not a
universal characteristic of NLS1s (Williams et al. \cite{williams04}). Wang \& Netzer 
(\cite{wang03}) presented a model consisting of an extreme slim disc with a hot corona 
to explain the soft X-ray excess and suggested that it is a natural consequence of super 
Eddington accretion. 

 The X-ray photon index of Mark\,110 in the energy range 0.2-2.0 keV ($\Gamma$=2.41$\pm$0.03, 
Lawrence et al. \cite{lawrence97} or $\Gamma$=2.47$\pm$0.01, Grupe et al. \cite{grupe01}) 
is typical of BLS1s rather than of NLS1s (Lawrence et al. \cite{lawrence97}) suggesting 
that this object is not super Eddington (Grupe \cite{grupe04}). Dasgupta \& Rao (\cite{dasgupta06}) found 
$\Gamma$=1.75$\pm$0.01 in the 2-12 keV range and a large soft excess which can be fitted 
with a blackbody with kT=100$\pm$2 eV. Alternatively, they could fit the data in the range 
0.3-12 keV with a broken power law, the values of the photon indices being 
$\Gamma$=2.29$\pm$0.01 and $\Gamma$=1.78$\pm$0.01 and the break energy 1.66$\pm$0.04 keV. 
This low value of the 2-12 keV photon index is again typical of BLS1s (Leighly \cite{leighly99}, 
Middleton et al. \cite{middleton07}). \\

 Xu et al. (\cite{xu07}) have shown that, while in NLS1s the electron density of the NLR, as 
estimated from the [S\,II]$\lambda$6716/$\lambda$6731 line ratio, covers a rather large range
(2-770 cm$^{-3}$, corresponding to $\lambda$6716/$\lambda$6731 in the range 0.94-1.23), in BLS1s 
the density is always relatively large ($>$140 cm$^{-3}$, $\lambda$6716/$\lambda$6731 $<$1.27).
In Mark\,110, we have measured $\lambda$6716/$\lambda$6731=1.12 which does not exclude the 
possibility that this object is an NLS1 but makes its classification as a BLS1 more likely. \\

 Fe\,II emission in the broad line region has not been detected and is extremely weak (see above).
Boroson (\cite{boroson02}) and Grupe (\cite{grupe04}) suggested that the inverse correlation 
between the strengths of Fe\,II and [O\,III] is driven predominantly by the Eddington ratio. 
Objects with a high Edington ratio have strong Fe\,II emission. The extreme weakness of Fe\,II 
in Mark\,110 therefore also argues for a low Eddington ratio. \\

 If Mark\,110 has a relatively low Eddington ratio, its BH mass should be larger than the 
typical value obtained for NLS1s of similar luminosity. 

 The Eddington luminosity is taken to be L$_{\rm Edd}$=1.25$\times$M$_{\rm BH}$/M$\odot$$\times$10$^{38}$
erg s$^{-1}$ (Laor et al. \cite{laor97}) {\it i.e.}, for Mark\,110, 17.4$\times$10$^{45}$ erg s$^{-1}$,
assuming M$_{\rm BH}$=14$\times$10$^{7}$ M$\odot$. The optical luminosity 
$\lambda$$\times$L$_{\lambda}$ at 5100 \AA\ is found to vary in the range 
(0.11-0.25)$\times$10$^{44}$ erg s$^{-1}$ 
(Peterson et al. \cite{peterson98}; Kaspi et al. \cite{kaspi05}) after removal of the host 
galaxy contribution (Bentz et al. \cite{bentz06}). In these conditions, the bolometric
luminosity varies in the range (0.11-0.25)$\times$10$^{45}$ erg s$^{-1}$ which corresponds to 
an Eddington ratio in the range (0.6-1.4)$\times$10$^{-2}$. If this is the case, Mark\,110 
would be far from emitting at the Eddington luminosity. Even if the BH mass is much smaller 
({\it e.g.} 0.33$\times$10$^{7}$ M$\odot$), the Eddington ratio would be in the range 0.25-0.60, 
still smaller than one.  

\subsection{Comparison of Mark\,110 with I\,Zw\,1 and IRAS\,07598+6508}

\begin{figure}[ht]

\resizebox{8.8cm}{!}{\includegraphics{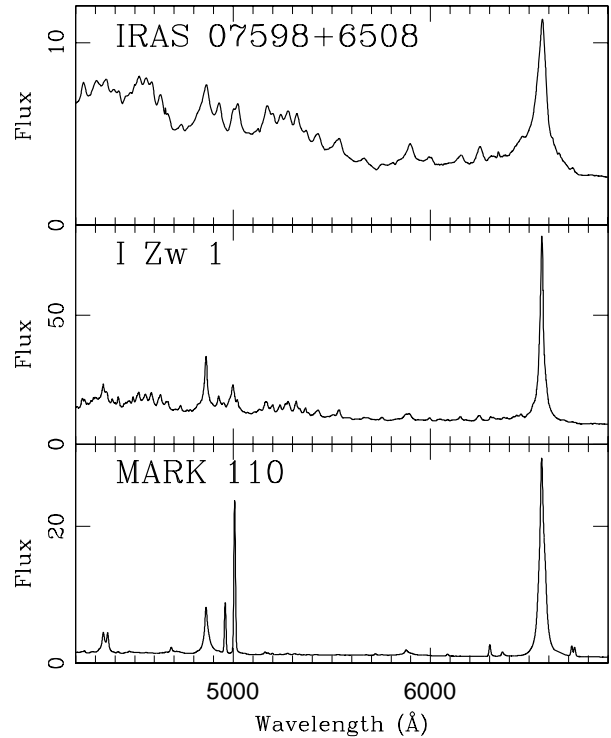}}

\caption{\label{nls1} Spectra of three NLS1s. Fluxes are in units of 10$^{-15}$
erg s$^{-1}$ cm$^{-2}$ \AA$^{-1}$. As shown in the text, although Mark\,110 has 
narrow "broad" emission lines, it is probably not a genuine NLS1.}
\end{figure}

 Mark\,110 is the third narrow line Seyfert 1 galaxy for which we have performed a detailed 
analysis of the emission line spectrum using high signal to noise spectra. The first two were 
I\,Zw\,1 (V\'eron-Cetty et al. \cite{veron04}) and IRAS\,07698+6508 (V\'eron-Cetty et al. 
\cite{cetty06}). These three objects have been classified as NLS1s on the basis of the width 
of their broad emission lines. Their spectra are extremely dissimilar (Fig.\,\ref{nls1}). 

 IRAS\,07598+6508 is a very strong Fe\,II emitter with R$_{\rm 4570}$=Fe\,II 
$\lambda$4570/H$\beta$$\sim$8. The spectrum is completely dominated by broad permitted 
lines of Fe\,II, Ti\,II and Cr\,II. No narrow line could be detected. In I\,Zw\,1, the broad 
permitted Fe\,II lines are weaker (R$_{\rm 4570}$=1.5-1.9). The narrow line system is relatively 
weak and is dominated by Fe\,II permitted and forbidden lines. In Mark\,110, no broad Fe\,II 
lines could be detected while the narrow line system is much stronger relative to the broad lines. 

 Vestergaard \& Peterson (\cite{vestergaard06}) have determined the luminosity of the nucleus 
and the BH mass of I\,Zw\,1. They found $\lambda$$\times$L(5100)= 6.22$\times$10$^{40}$ erg s$^{-1}$
and M$_{\rm BH}$=2.65$\times$10$^{7}$ M$\odot$ which leads to an Eddington ratio of 1.8. 

 The flux density of IRAS\,07598+6508 at 5100 \AA\ is 3.64$\times$10$^{-16}$ erg s$^{-1}$ cm$^{-2}$
\AA$^{-1}$ which, with z= 0.149, corresponds to
$\lambda$$\times$L(5100)=8.9$\times$10$^{44}$ erg s$^{-1}$. According to Bentz et al. 
(\cite{bentz06}) this corresponds to a size of the broad line region of 100 l.d. Then, according to 
Kaspi et al. (\cite{kaspi05}) and using k=1.26 and 1\,780 km s$^{-1}$ for the H$\beta$ FWHM 
(V\'eron-Cetty et al. \cite{veron01}), the BH mass is M$_{\rm BH}$=7.34$\times$10$^{7}$ M$\odot$.
We deduced from these numbers that the Eddington ratio is equal to $\sim$1. \\

 The values of the Eddington ratio for these three galaxies are quite uncertain but seem 
to confirm that this is an important parameter in determining the main properties of the emission 
line spectrum of this type of object.

\subsection{Other similar objects}

 Kaz 320 is a Seyfert 1 galaxy (B=13.8) at z=0.034. The FWHM of the broad H$\beta$ component 
is equal to 1\,375 km s$^{-1}$ (Botte et al. \cite{botte04}) or 1\,470 km s$^{-1}$ (V\'eron-Cetty 
et al. \cite{veron01}). The spectrum exhibits strong Balmer and [O\,III] lines together with 
some highly ionized species like [Fe\,VII]$\lambda$6087 and [Fe\,X]$\lambda$6375. According 
to Zamorano et al. (\cite{zamorano92}), permitted Fe\,II lines, if present, are too weak to 
be detected. However we have measured R$_{4570}$=0.49. We found that the H$\beta$ FWHM of the 
broad component is equal to 1\,470 km s$^{-1}$ (V\'eron-Cetty et al. \cite{veron01}).

 HS\,0328+0528 is a Seyfert 1 galaxy (B=15.7) at z=0.043 (Perlman et al. \cite{perlman96}; 
Engels et al. \cite{engels98}). The FWHM of the broad Balmer component is  
$\sim$1\,500 km s$^{-1}$. The Fe\,II emission is weak (R$_{4570}$=0.43)(V\'eron-Cetty et 
al. \cite{veron01}). \\

 The BH mass has been estimated to be 0.23 and 0.53$\times$10$^{7}$ M$\odot$ respectively in 
these two objects (Wang \& Lu \cite{wang01}).
Like Mark\,110, they could be pole-on, but otherwise normal BLS1s.

\section{Conclusion}

 We have analyzed the optical emission line spectrum of the peculiar NLS1 galaxy Mark\,110. 
Except for "narrow" broad lines, Mark\,110 lacks all other characteristics of this class of 
objects such as a weak [O III]$\lambda5007$/H$\beta$ ratio and strong permitted Fe II lines
(We have shown that all the detected Fe II lines belong to the narrow line system). The X-ray 
spectrum of Mark\,110 is also more similar to that of a BLS1 than that of an NLS1.

 The properties of the NLS1s are generally attributed to a high accretion rate on a relatively 
small mass BH leading to a super Eddington luminosity. We argue that, although the "broad" 
emission lines in Mark\,110 are narrow ($\sim$1\,700 km s$^{-1}$), its BH mass is such that 
its luminosity is not super Eddington.

 The analysis of the narrow line system indicates that the electron density fills the range 
10$^{3}$ - 10$^{6}$ cm$^{-3}$ with a column density between 5$\times$10$^{19}$ and 
5$\times$10$^{21}$ cm$^{-2}$.

 The broad lines have three components with a FWHM of 6\,000, 3\,340 and 1\,515 km s$^{-1}$ 
respectively. The first and the third components are variable while, strangely, the second is not.

 Comparison with two previously studied NLS1s (IRAS 07598+6508 and I Zw 1) shows the diversity of 
the emission spectrum of these objects which exhibit a very large range of Fe II emission intensity.
The total width of the broad lines, although easy to measure, is probably not a major significant 
physical parameter to classify these objects. The Eddington luminosity, much more difficult to 
evaluate, is certainly more important.

 We note that a few other objects such as Kaz\,320 and HS\,0328+0528, are similar to Mark\,110 
in having "narrow" broad lines, but weak Fe II and strong [O\,III], constituting a sub-class of 
BLS1s with "narrow" broad lines.

\section{Acknowledgments}

 We gratefully thank M.C. Bentz who kindly put at our disposal her HST image 
of Mark\,110 and S. Collin and D. P\'equignot for helpful discussions. We acknowledge the referee, D. Grupe, 
thanks to whom the paper has been significantly improved. 

{}
 \appendix
\section{Line list for the narrow line region}

\begin{table}[t]
\caption{\label{narrow}Lines observed in the stronger narrow line system. 
Col. 1 : line identification, col. 2 : rest wavelength, col. 3 : intensity relative to 
H$\beta$ (H$\beta$ flux = 20.3$\times$10$^{-15}$ erg s$^{-1}$ cm$^{-2}$).}
\begin{center} 
\begin{tabular}{|l|l|r|}
\hline 
Name & wavelength & intensity  \\ 
\hline 
 H$\beta$           & 4861.33 &  1.00  \\
 H$\alpha$          & 6562.77 &  3.02  \\
 H$\gamma$          & 4340.47 &  0.48  \\
 Ti II 33           & 4227.34 &  0.04  \\
 $\rm [Fe II]$ 21F  & 4231.56 &  0.00  \\ 
 Fe II 27           & 4233.17 &  0.02  \\
 Cr II 31           & 4233.25 &  0.02  \\
 Cr II 17           & 4238.69 &  0.03  \\
 $\rm [Fe II]$ 21F  & 4243.97 &  0.10  \\ 
 $\rm [Fe II]$ 21F  & 4244.81 &  0.02  \\ 
 $\rm [Fe II]$ 21F  & 4276.83 &  0.06  \\ 
 $\rm [Fe II]$  7F  & 4287.39 &  0.09  \\ 
 $\rm [Fe II]$ 21F  & 4305.90 &  0.02  \\ 
 $\rm [Fe II]$ 21F  & 4319.63 &  0.04  \\ 
 He II  3           & 4338.67 &  0.01  \\
 $\rm [Fe II]$ 21F  & 4346.85 &  0.02  \\ 
 $\rm [Fe II]$ 21F  & 4352.78 &  0.03  \\ 
 $\rm [Fe II]$ 21F  & 4358.36 &  0.04  \\ 
 $\rm [Fe II]$  7F  & 4359.33 &  0.07  \\ 
 $\rm [O III]$  2F  & 4363.21 &  0.77  \\ 
 Ti II 104          & 4367.66 &  0.10  \\
 $\rm [Fe II]$ 21F  & 4372.43 &  0.02  \\ 
 Ti II 93           & 4374.82 &  0.04  \\
 He I 51            & 4387.93 &  0.00  \\
 $\rm [Fe II]$  7F  & 4413.78 &  0.05  \\ 
 $\rm [Fe II]$  6F  & 4416.27 &  0.05  \\ 
 $\rm [Fe II]$  6F  & 4432.45 &  0.00  \\ 
 Ti II 19           & 4443.80 &  0.01  \\
 $\rm [Fe II]$  7F  & 4452.10 &  0.03  \\ 
 $\rm [Fe II]$  6F  & 4457.94 &  0.02  \\ 
 He I 14            & 4471.69 &  0.05  \\
 $\rm [Fe II]$  7F  & 4474.90 &  0.02  \\ 
 Mg II  4           & 4481.13 &  0.03  \\
 Ti II 115          & 4488.32 &  0.02  \\
 $\rm [Fe II]$  6F  & 4488.75 &  0.01  \\ 
 Fe II 37           & 4489.18 &  0.01  \\
 Fe II 37           & 4491.40 &  0.03  \\
 $\rm [Fe II]$  6F  & 4492.63 &  0.01  \\ 
 Ti II 31           & 4501.27 &  0.06  \\
 Fe II 38           & 4508.28 &  0.05  \\
 $\rm [Fe II]$  6F  & 4509.60 &  0.00  \\ 
 $\rm [Fe II]$  6F  & 4514.90 &  0.01  \\ 
 Fe II 37           & 4515.34 &  0.05  \\
 Fe II 37           & 4520.22 &  0.02  \\
 Fe II 38           & 4522.63 &  0.03  \\
 $\rm [Fe II]$  6F  & 4528.38 &  0.00  \\ 
 Ti II 82           & 4529.46 &  0.02  \\
 $\rm [Fe II]$  6F  & 4533.00 &  0.00  \\ 
 He II  2           & 4541.49 &  0.05  \\
 Fe II 38           & 4549.47 &  0.04  \\
 Fe II 37           & 4555.89 &  0.02  \\
 Cr II 44           & 4558.68 &  0.03  \\
 Ti II 50           & 4563.76 &  0.03  \\
 Ti II 81           & 4571.97 &  0.03  \\
 Fe II 38           & 4576.33 &  0.03  \\
 Fe II 37           & 4582.83 &  0.01  \\

\hline
\end{tabular}
\end{center}
\end{table}
\addtocounter{table}{-1}
\begin{table}[ht]
\caption{(continued)} 
\begin{center}
\begin{tabular}{|l|l|r|}
\hline 
Name & wavelength & intensity  \\ 
\hline

 Fe II 38           & 4583.83 &  0.04  \\
 Ti II 50           & 4589.96 &  0.04  \\
 Fe II 38           & 4620.51 &  0.00  \\
 Fe II 37           & 4629.34 &  0.03  \\
 $\rm [Fe II]$  4F  & 4639.67 &  0.00  \\ 
 ?                  & 4642.78 &  0.02  \\
 $\rm [Fe III]$  3F & 4658.05 &  0.05  \\ 
 $\rm [Fe II]$  4F  & 4664.44 &  0.00  \\ 
 He II  1           & 4685.68 &  0.22  \\
 $\rm [Ar IV]$  1F  & 4711.37 &  0.01  \\ 
 He I 12            & 4713.17 &  0.00  \\
 $\rm [Ne IV]$  1F  & 4714.25 &  0.02  \\ 
 $\rm [Ne IV]$  1F  & 4715.61 &  0.01  \\ 
 $\rm [Ne IV]$  1F  & 4724.15 &  0.02  \\ 
 $\rm [Ne IV]$  1F  & 4725.62 &  0.02  \\ 
 $\rm [Fe II]$  4F  & 4728.07 &  0.01  \\ 
 Fe II] 43          & 4731.45 &  0.02  \\
 $\rm [Ar IV]$  1F  & 4740.16 &  0.04  \\ 
 $\rm [Fe II]$  4F  & 4772.06 &  0.00  \\ 
 $\rm [Fe II]$ 20F  & 4774.72 &  0.01  \\ 
 $\rm [Fe II]$  4F  & 4798.27 &  0.00  \\ 
 $\rm [Fe II]$ 20F  & 4814.53 &  0.05  \\ 
 Cr II 30           & 4824.13 &  0.02  \\
 $\rm [Fe II]$ 20F  & 4874.48 &  0.01  \\ 
 $\rm [Fe II]$ 20F  & 4905.34 &  0.02  \\ 
 He I 48            & 4921.93 &  0.01  \\
 Fe II 42           & 4923.92 &  0.05  \\
 $\rm [Fe III]$  1F & 4930.53 &  0.05  \\ 
 $\rm [Fe II]$ 20F  & 4947.38 &  0.01  \\ 
 $\rm [Fe II]$ 20F  & 4950.74 &  0.01  \\ 
 $\rm [O III]$  1F  & 4958.91 &  2.98  \\ 
 $\rm [Fe II]$ 20F  & 4973.39 &  0.01  \\ 
 $\rm [Fe II]$ 20F  & 5005.51 &  0.01  \\ 
 $\rm [Fe II]$  4F  & 5006.65 &  0.00  \\ 
 $\rm [O III]$  1F  & 5006.84 &  8.96  \\ 
 He I  4            & 5015.67 &  0.02  \\
 Fe II 42           & 5018.43 &  0.08  \\
 $\rm [Fe II]$ 20F  & 5020.23 &  0.01  \\ 
 $\rm [Fe II]$ 20F  & 5043.52 &  0.01  \\ 
 $\rm [Fe II]$ 18F  & 5107.94 &  0.00  \\ 
 $\rm [Fe II]$ 19F  & 5111.63 &  0.01  \\ 
 $\rm [Fe II]$ 18F  & 5158.00 &  0.01  \\ 
 $\rm [Fe II]$ 19F  & 5158.78 &  0.07  \\ 
 $\rm [Fe II]$ 35F  & 5163.95 &  0.05  \\ 
 Fe II 42           & 5169.03 &  0.03  \\
 $\rm [Fe II]$ 18F  & 5181.95 &  0.00  \\ 
 $\rm [Ar III]$  3F & 5191.82 &  0.03  \\ 
 Fe II 49           & 5197.57 &  0.01  \\
 $\rm [N I]$  1F    & 5197.90 &  0.01  \\ 
 $\rm [Fe II]$ 35F  & 5199.17 &  0.01  \\ 
 $\rm [N I]$  1F    & 5200.26 &  0.04  \\ 
 $\rm [Fe II]$ 19F  & 5220.06 &  0.01  \\ 
 Fe II 49           & 5234.62 &  0.03  \\
 $\rm [Fe II]$ 19F  & 5261.62 &  0.04  \\ 
 $\rm [Fe II]$ 18F  & 5268.88 &  0.00  \\ 
 $\rm [Fe III]$ 1F  & 5270.40 &  0.03  \\ 
 $\rm [Fe II]$ 18F  & 5273.35 &  0.01  \\ 
 Fe II 49           & 5275.99 &  0.03  \\

\hline
\end{tabular}
\end{center}
\end{table}
\addtocounter{table}{-1}
\begin{table}[ht]
\caption{(end)} 
\begin{center}
\begin{tabular}{|l|l|r|}
\hline 
Name & wavelength & intensity  \\ 
\hline

 $\rm [Fe II]$ 35F  & 5278.37 &  0.01  \\ 
 $\rm [Fe II]$ 35F  & 5283.11 &  0.01  \\ 
 ?                  & 5288.90 &  0.02  \\
 $\rm [Fe II]$ 19F  & 5296.83 &  0.01  \\ 
 Fe II 49           & 5316.61 &  0.04  \\
 Fe II 49           & 5325.56 &  0.02  \\
 $\rm [Fe II]$ 19F  & 5333.65 &  0.03  \\ 
 Ti II 69           & 5336.81 &  0.02  \\
 $\rm [Fe II]$ 18F  & 5347.65 &  0.00  \\ 
 $\rm [Fe II]$ 19F  & 5376.47 &  0.02  \\ 
 He II  2           & 5411.52 &  0.00  \\
 $\rm [Fe II]$ 17F  & 5412.65 &  0.03  \\ 
 Fe II 49           & 5425.27 &  0.02  \\
 $\rm [Fe II]$ 18F  & 5433.13 &  0.00  \\ 
 $\rm [Fe II]$ 34F  & 5477.25 &  0.01  \\ 
 $\rm [Fe II]$ 17F  & 5495.82 &  0.01  \\ 
 $\rm [Fe II]$ 17F  & 5527.34 &  0.04  \\ 
 Fe II] 55          & 5534.86 &  0.03  \\
 $\rm [Fe II]$ 18F  & 5556.31 &  0.00  \\ 
 $\rm [O I]$  3F    & 5577.34 &  0.01  \\ 
 ?                  & 5608.74 &  0.01  \\
 $\rm [Fe II]$ 17F  & 5654.85 &  0.01  \\ 
 $\rm [Fe II]$ 17F  & 5745.70 &  0.00  \\ 
 $\rm [Fe II]$ 34F  & 5746.97 &  0.01  \\ 
 $\rm [N II]$   3F  & 5754.57 &  0.03  \\ 
 $\rm [Fe IV]$      & 5798.78 &  0.01  \\ 
 $\rm [Fe II]$ 34F  & 5843.90 &  0.00  \\ 
 He I 11            & 5875.70 &  0.10  \\
 Na I D             & 5889.95 &  0.03  \\
 Na I D             & 5895.92 &  0.03  \\
 $\rm [O I]$  1F    & 6300.23 &  0.60  \\ 
 $\rm [S III]$  1F  & 6312.06 &  0.07  \\ 
 Si II  2           & 6347.09 &  0.02  \\
 $\rm [O I]$  1F    & 6363.88 &  0.20  \\ 
 Si II  2           & 6371.36 &  0.05  \\
 $\rm [N II]$  1F   & 6548.04 &  0.14  \\ 
 He II  2           & 6560.10 &  0.02  \\
 $\rm [N II]$  1F   & 6583.46 &  0.43  \\ 
 He I 46            & 6678.15 &  0.03  \\
 $\rm [S II]$  2F   & 6716.44 &  0.58  \\ 
 $\rm [S II]$  2F   & 6730.81 &  0.51  \\ 
 $\rm [Fe VII]$  2F & 4893.90 &  0.01  \\ 
 $\rm [Ca VII]$  1F & 4940.30 &  0.02  \\ 
 $\rm [Fe VII]$  2F & 4942.49 &  0.04  \\ 
 $\rm [Fe VI]$  2F  & 4967.13 &  0.06  \\ 
 $\rm [Fe VI]$  2F  & 4972.47 &  0.09  \\ 
 $\rm [Fe VI]$  2F  & 5145.75 &  0.03  \\ 
 $\rm [Fe VII]$  2F & 5158.41 &  0.09  \\ 
 $\rm [Fe VI]$  2F  & 5176.04 &  0.14  \\ 
 $\rm [Fe VII]$  2F & 5276.39 &  0.05  \\ 
 $\rm [Fe XIV]$  1F & 5303.60 &  0.01  \\ 
 $\rm [Ca V]$  1F   & 5308.90 &  0.08  \\ 
 $\rm [Ca VII]$  1F & 5620.36 &  0.01  \\ 
 $\rm [Fe VI]$  1F  & 5631.07 &  0.01  \\ 
 $\rm [Ca VI]$  2F  & 5631.40 &  0.01  \\ 
 $\rm [Fe VI]$  1F  & 5676.95 &  0.03  \\ 
 $\rm [Fe VII]$  1F & 5720.71 &  0.13  \\ 
 $\rm [Fe VII]$  1F & 6086.30 &  0.18  \\ 
\hline
\end{tabular}
\end{center}
\end{table}


\begin{thebibliography}{}

 \bibitem[1977]{adams77} 
 Adams, T.F. 1977, ApJS, 33,19 
 \bibitem[1988]{appenzeller88}
 Appenzeller, I. \& Ostreicher, R. 1988, AJ, 95,45
 \bibitem[1970]{arakelyan70}
 Arakelyan, M.A., Dibai, E.A., Esipov, V.F. \& Markaryan, B.E. 1970, Astrophysics, 6,189 
 \bibitem[2005]{barth05}
 Barth, A.J., Greene, J.E. \& Ho, L.C. 2005, ApJ, 619,L151
 \bibitem[2005]{baskin05}
 Baskin, A. \& Laor, A. 2005, MNRAS, 358,1043 
 \bibitem[2006]{bentz06}
 Bentz, M.C., Peterson, B.M., Pogge, R.W., Vestergaard, M. \&  Onken, C.A. 2006, ApJ, 644,133
 \bibitem[1999]{bischoff99}
 Bischoff, K. \& Kollatschny, W. 1999, A\&A, 345,49 
 \bibitem[2007]{boller07}
 Boller, T., Balestra, I. \& Kollatschny, W. 2007, A\&A, 465,87
 \bibitem[2002]{boroson02}
 Boroson, T.A. 2002, ApJ, 565,78
 \bibitem[1992]{boroson92}
 Boroson, T.A. \& Green, R.F. 1992, ApJS, 80,109
 \bibitem[2004]{botte04}
 Botte, V., Ciroi, S., Rafanelli, P. \& Di Mille, F. 2004, AJ, 127,3168
 \bibitem[2005]{botte05}
 Botte, V., Ciroi, S., Di Mille F., Rafanelli, P. \& Romano, A. 2005, MNRAS, 356,789
 \bibitem[2000]{collin00}
 Collin, S. \& Joly, M. 2000, New Astronomy Review, 44,531
 \bibitem[2004]{collin04}
 Collin, S. \& Kawaguchi, T. 2004, A\&A, 426,797
 \bibitem[2005]{cool05}
 Cool, R.J., Howell, S.B., Pena, M., Adamson, A.J. \& Thompson, R.R. 2005, PASP, 117,462
 \bibitem[1999]{crawford99}
 Crawford, F.L., McKenna, F.C., Keenan, F.P. et al. 1999, A\&AS, 139,135
 \bibitem[1986]{crenshaw86}
 Crenshaw, D.M. 1986, ApJS, 62,821
 \bibitem[2006]{dasgupta06}
 Dasgupta, S. \& Rao, A.R. 2006, ApJ, 651,L13
 \bibitem[1984]{robertis84}
 De Robertis, M.M. \& Osterbrock, D.E. 1984, ApJ, 286,171
 \bibitem[1999]{elvis99}
 Elvis, A., Wilkes, B.J., McDowell, J.C. et al. 1999, ApJS, 95,1
 \bibitem[1998]{engels98}
 Engels, D., Hagen, H.-J., Cordis, L. et al. 1998, A\&AS, 128,507
 \bibitem[1982]{feldman82}
 Feldman, F.R., Weedman, D.W., Balzano, V.A. \& Ramsey, L.W. 1982, ApJ, 256,427
 \bibitem[2000]{ferrarese00}
 Ferrarese, L. \& Merritt, D. 2000, ApJ, 539,L9 
 \bibitem[2001]{ferrarese01}
 Ferrarese, L., Pogge, R.W., Peterson, B.M. et al. 2001, ApJ, 555,L79
 \bibitem[1999]{lario99}
 Garcia-Lario, P., Riera, A. \& Manchado, A. 1999, ApJ, 526,854
 \bibitem[2006]{greene06}
 Greene, J.E. \& Ho, L.C. 2006, ApJ, 641,L21
 \bibitem[2004]{grupe04} 
 Grupe, D. 2004, AJ, 127,1799 
 \bibitem[1999]{grupe99} 
 Grupe, D., Beuermann, K., Mannheim, K. \& Thomas, H.-C. 1999, A\&A, 350,805
 \bibitem[2001]{grupe01}
 Grupe, D., Thomas, H.-C. \& Beurmann, K. 2001, A\&A, 367,470
 \bibitem[2004]{grupe04b} 
 Grupe D., Wills, B.J., Leighly, K.L. \& Meusinger, H. 2004, AJ, 127,156
 \bibitem[1999]{harlaftis99}
 Harlaftis, E. 1999, A\&A, 346,L73
 \bibitem[1988]{hutchings88}
 Hutchings, J.B. \& Craven, S.E. 1988, AJ, 95,677
 \bibitem[2000]{kaspi00}
 Kaspi, S., Smith, P.S., Netzer, H. et al. 2000, ApJ, 533,631
 \bibitem[2005]{kaspi05}
 Kaspi, S., Maoz, D., Netzer, H. et al. 2005, ApJ, 629,61
 \bibitem[2001]{keenan01}
 Keenan, F.P., Crawford, F.L., Feibelman, W.A. \& Aller, L.H. 2001, ApJS, 132,103
 \bibitem[2003a]{kollatschny03a}
 Kollatschny, W. 2003a, A\&A, 407,461
 \bibitem[2003b]{kollatschny03b}
 Kollatschny, W. 2003b, A\&A, 412,L61
 \bibitem[1981]{kollatschny81}
 Kollatschny, W. \& Fricke, K.J. 1981, A\&A, 146,L11
 \bibitem[1981]{kollatschny81a}
 Kollatschny, W., Schneider, H., Fricke, K.J. \& York, H.W. 1981, A\&A, 104,198
 \bibitem[2000]{kollatschny00}
 Kollatschny, W., Bischoff, K. \& Dietrich, M. 2000, A\&A, 361,901
 \bibitem[2001]{kollatschny01}
 Kollatschny, W., Bischoff, K., Robinson, E.L., Welsh, W.F. \& Hill, G.J. 2001, A\&A, 379,125 
 \bibitem[2001]{krolik01}
 Krolik, J.H. 2001, ApJ, 551,72
 \bibitem[1997]{laor97}
 Laor, A., Fiore, F., Elvis, M., Wilkes, B.J. \& McDowel, J.C. 1997, ApJ, 477,93
 \bibitem[1997]{lawrence97}
 Lawrence, A., Elvis, M., Wilkes, B.J., McHardy, I. \& Brandt, N. 1997, MNRAS, 285,879
 \bibitem[1999]{leighly99}
 Leighly, K.M. 1999, ApJS, 125,317
 \bibitem[1990]{mckenty90}
 McKenty, J.W. 1990, ApJS, 72,231 
 \bibitem[1969]{markaryan69}
 Markaryan, B.E. 1969, Astrophysics, 5,206
 \bibitem[1997]{mckenna97}
 McKenna, F.C., Keenan, F.P., Hambly, N.C. et al. 1997, ApJS, 109,225
 \bibitem[1961]{merrill61}
 Merrill, P. W. 1961, ApJ, 133,503
 \bibitem[2001]{merritt01}
 Merritt, D. \& Ferrarese, L. 2001, ApJ, 547,140
 \bibitem[1985]{meyers85}
 Meyers, K.A. \& Peterson, B.M. 1985, PASP, 97,734
 \bibitem[2007]{middleton07} 
 Middleton, M., Done, C. \& Gierlinski, M. 2007, MNRAS (submitted), astro-ph/0704.2970 
 \bibitem[2006]{muller06} 
 M\"{u}ller, A. \& Wold, M. 2006, A\&A, 457,485
 \bibitem[1990]{netzer90}
 Netzer, H. 1990, in: Active galactic nuclei, R.D. Blanford, Netzer, H. \& Woltjer, L. eds.
 \bibitem[2005]{oneill05}
 O'Neill, P.M., Nandra, K., Papadakis, I.E. \& Turner, T.J. 2005, MNRAS, 358,1405
 \bibitem[2004]{onken04}
 Onken, A., Ferrarese, L., Merritt, D. et al. 2004, ApJ, 615,645
 \bibitem[1974]{osterbrock74}
 Osterbrock, D.E. 1974, Astrophysics of gaseous nebulae, W.H. Freeman and company 
 \bibitem[1977]{osterbrock77}
 Osterbrock, D.E. 1977, ApJ, 215,733
 \bibitem[1985]{osterbrock85}
 Osterbrock, D.E. \& Pogge, R.W. 1985, ApJ, 297,166
 \bibitem[2004]{papadakis04}
 Papadakis, I.E. 2004, MNRAS, 348,207
 \bibitem[1996]{perlman96}
 Perlman, E.S., Stocke, J.T., Schachter, J.F. et al. 1996, ApJS, 104,251 
 \bibitem[1988]{peterson88}
 Peterson, B.M. 1988, PASP, 100,18
 \bibitem[2000]{peterson00}
 Peterson, B.M. \& Wandel, A. 2000, ApJ, 540,L13
 \bibitem[1984]{peterson84}
 Peterson, B.M., Foltz, C.B., Crenshaw, D.M., Meyers, K.A. \& Byard, M.C. 1984, ApJ, 279,529 
 \bibitem[1985]{peterson85}
 Peterson, B.M., Crenshaw, D.M. \& Meyers, K.A. 1985, ApJ, 298,283
 \bibitem[1998]{peterson98}
 Peterson, B.M., Wanders, I., Bertram, R. et al. 1998, ApJ, 501,82 (erratum in ApJ, 511,513)
 \bibitem[2004]{peterson04}
 Peterson, B.M., Ferrarese, L., Gilbert, K.M. et al. 2004, ApJ, 613,682  
 \bibitem[1978]{phillips78}
 Phillips, M.M. 1978, ApJ, 226,736
 \bibitem[1995]{pounds95}
 Pounds, K.A., Done, C. \& Osborne, J.P. 1995, MNRAS, 277,L5
 \bibitem[1998]{schlegel98}
 Schlegel, D.J., Finkbeiner, D.P. \& Davis, M. 1998, ApJ, 500,525
 \bibitem[2006]{shemmer06}
 Shemmer, O., Brandt, W.N., Netzer, H., Maiolino, R. \& Kaspi, S. 2006, ApJ, 646,L29
 \bibitem[2003]{stepanian03}
 Stepanian, J.A., Benitez, E., Krongold, Y. et al. 2003, ApJ, 588,746
 \bibitem[2000]{storey00}
 Storey, P.J. \& Zeippen, C.J. 2000, MNRAS, 312,813 
 \bibitem[1980]{veron80}
 V\'eron, P., Lindblad, P.O., Zuiderwijk, E.J., V\'eron-Cetty, M.-P. \& Adams, G.
  1980, A\&A, 87,245
 \bibitem[2001]{veron01}
 V\'eron-Cetty, M.-P., V\'eron, P. \& Goncalves, A.C. 2001, A\&A, 372,730
 \bibitem[2004]{veron04}
 V\'eron-Cetty, M.-P., Joly, M. \& V\'eron, P. 2004, A\&A, 417,515
 \bibitem[2006]{cetty06}
 V\'eron-Cetty, M.-P., Joly, M. \& V\'eron, P. 2006, A\&A, 451,851
 \bibitem[2006]{vestergaard06}
 Vestergaard, M. \& Peterson, B.M. 2006, ApJ, 625,688
 \bibitem[1985]{vrtilek85}
 Vrtilek, J.M. \& Carleton, N.P. 1985, ApJ, 294,106
 \bibitem[1999]{wandel99}
 Wandel, A., Peterson, B.M. \& Malkan, M.A. 1999, ApJ, 526,579
 \bibitem[2001]{wang01}
 Wang, T. \& Lu, Y. 2001, A\&A, 377,52
 \bibitem[2003]{wang03}
 Wang, J.M. \& Netzer, H. 2003, A\&A, 398,927
 \bibitem[2005]{wang05}
 Wang, J., Wei, J.Y. \& He, X.T. 2005, A\&A, 436,417
 \bibitem[2000]{webb00}
 Webb, W. \& Malkan, M. 2000, ApJ, 540,652
 \bibitem[1977]{wehinger77}
 Wehinger, P.A. \& Wyckoff, S. 1977, MNRAS, 72,231
 \bibitem[2004]{williams04}
 Williams, R.J., Mathur, S. \& Pogge, R.W. 2004, ApJ, 610,737
 \bibitem[2003]{wu03}
 Wu, J.-H., He, X.-T., Chen, Y. \& Voges, W. 2003, Chin. J. Astron. Astrophys., 3,423
 \bibitem[2007]{xu07} 
 Xu, D., Komossa, S., Zhou, H., Wang, T., Wei, J. 2007,ApJ (in press), astro-ph/0706.2574
 \bibitem[1992]{zamorano92}
 Zamorano, J., Gallego, J., Rego, M., Vitores, A.G. \& Gonzalez-Riestra, R. 1992, AJ, 104,1000
 \end{thebibliography}
 \end{document}